\documentclass[
 reprint,
 amsmath,amssymb,
 aps
]{revtex4-2}
\usepackage{xcolor}
\usepackage{graphicx}% Include figure files
\usepackage{dcolumn}% Align table columns on decimal point
\usepackage{bm}% bold math
\begin{document}

\title{Femtoscopy analysis of ultrasoft pion trap at energies available at the CERN Large Hadron Collider}
%{Femtocopic study of ultra-soft pions  at LHC energies}
\author{W. Rzesa$^{1}$, G. Kornakov$^{1}$, A.R. Kisiel$^{1}$, Yu.M. Sinyukov$^{1,2}$, V.M. Shapoval$^{2}$}
\affiliation{$^{1}$ Warsaw University of Technology, Faculty of Physics, ul. Koszykowa 75, 00-662, Warsaw, Poland\\
 $^{2}$ Bogolyubov Institute for Theoretical Physics, 14b Metrolohichna street, Kyiv 03143, Ukraine 
}

\date{\today}% It is always \today, today,
             %  but any date may be explicitly specified

\begin{abstract}
Femtoscopy studies of pion radiation in heavy-ion collisions have been conducted extensively at all available collider energies, both theoretically and experimentally. 
%The femtoscopic studies of sizes of homogeneity region of particles coming from particles emitting source were subject of many investigations  at both experimental and theoretical levels. 
In all these studies a special interest is given to $m_{T}$ dependency of pion femtoscopy radii, usually approximated %described 
by a power-law function at transverse momenta above 200~MeV/$c$. 
However, the radii behaviour has been much less explored for the ultra-soft pions, %which have 
possessing the transverse momentum comparable to or lower than the pion mass. 
%never been experimentally and theoretically confirmed in very low $p_T/m_T$ region. 
%This region that so far has not been accessed due to detection limitation of detectors. 
For many experimental setups this region is difficult to measure.
In this work we present theoretical calculations of pion emission in the ultra-soft region in %hydrodynamical 
the two hybrid models --- iHKM and LHYQUID+THERMINATOR2. Along with the particle transverse momentum spectra, we present the calculated femtoscopy radii, both in one-dimensional and three-dimensional representations. We investigate the radii dependence 
on pair $m_{T}$ and observe, in particular, a departure from the power-law behaviour at ultra-soft momenta, potentially %signalizing about certain 
reflecting a decoupling of such slow pions from the rest of collectively expanding system.
%, where pion velocity becomes comparable to or lower than the collective velocity of the system expansion. 
We provide the theoretical interpretation of this result and discuss its significance, in particular, for the ongoing non-identical particle femtoscopy %investigations 
analysis for pairs consisting of a pion and a baryon (or of a pion and a charmed meson).
%This work concerns simulation studies of pion's femtoscopy and femtoscopic sizes measurement that cover very small values of $p_T$. The study has been done via 1- and 3-dimensional femtoscopy, for the two models (iHKM, LHYQUID+THERMINATOR2), 6 centrality intervals and 10 $m_T$ bins. This paper shows that the power-law dependence is definitely questionable for 1D studies and transverse directions of 3-dimensional studies in low $m_T$. 
\end{abstract}

\maketitle
%%%%%%%%%%%%%%%%%%%%%%%%%%%%%%%%%%%%%%%%%%%%%%%%%%%%%%%%%%%%%
%%%%%%%%%%%%%%%%introduction%%%%%%%%%%%%%%%%%%%%%%%%%%%%%%%%%
%%%%%%%%%%%%%%%%%%%%%%%%%%%%%%%%%%%%%%%%%%%%%%%%%%%%%%%%%%%%%

\section{\label{sec:level0}INTRODUCTION}
The interferometry of particles emitted from the small radiating source enables one to study the evolving geometry of that source. The particle interferometry has originated from the experimental paper~\cite{Goldhaber1960} and theoretical works~\cite{KOPYLOV1972, KOPYLOV1974}. In the latter it was also found that the results for the  quantum statistics correlation effect for two close sources of identical particle emission in collision physics are formally 
%(in formulas) 
similar to those for classical intensity correlations of the light radiation coming from different parts of the star, known in astronomy as the Hanbury Brown – Twiss (HBT) effect~\cite{hbt1954, hbt1956, hbt1957}. The similarity, as well as the difference between the quantum correlations in identical-particle pairs and the classical HBT effect are analyzed, e.g., in~\cite{sinyukov}. Note, that the correlation femtoscopy method, initially developed for finding the overall {\it size} of identical quantum particles' source, now is considered to measure the {\it homogeneity lengths}~\cite{sinyukovexpanding} of such femto-small sources. 
%(only in particular cases coinciding with the full system size). 
This general interpretation is especially important for the studies of the space-time structure of expanding sources, such as those created in 
%processes of
proton/antiproton and heavy-ion collisions~\cite{Goldhaber1960, KOPYLOV1972, KOPYLOV1974, KOONIN197743, Gyulassy1979, Zajc1984, Makhlin1986, Makhlin1988, Gersch1987}. 

%One of the main outcomes of the femtoscopic correlation studies is the measurements of the homogeneity region of the system. 
The homogeneity lengths, defined as a result of correlation femtoscopy analysis, reflect the spatial dimensions of the region 
%should not be misunderstood as the overall source of particles but as an area 
within the entire strongly interacting system formed in the collision, 
%full source 
from which particles are emitted with similar velocities
(having close, nearly coinciding
%, nearly {\it fixed} 
values and directions)~\cite{sinyukovexpanding, sinyukovqgs}. 
%One usually uses 
%The particle emitting source is usually referred the 
Gaussian parameterization for the particle emission source function (SF) and the two-particle correlation function (CF) are usually used in femtoscopy studies, although the realistic source shape differs from a perfect Gaussian --- e.g., resonance decay contributions cause the exponential behaviour near the peak of the correlation function. The conventional use of %same
the Gaussian shape allows one to standardize the description of experimental data and %make comparisons 
easily compare the results of different femtoscopy %radii 
measurements, as well as %allow their 
to interpret the obtained radii as the homogeneity lengths.
It is also well-motivated experimentally in heavy-ion collisions, where correlation shapes for pions in 3D analyses are universally observed to be Gaussian in a wide range of centralities and pair transverse momenta.
%It is important to note that 
The study of the dependence of the radii on the particle species, collision type and collision energy is the main objective of the femtoscopic analysis of heavy-ion collisions. 

In experimental studies, one clearly observes %well-defined 
certain types of universal scaling behaviour for the measured femtoscopy scales. %For example, 
For a given colliding system and collision energy a linear scaling of the femtoscopic radii is universally observed versus cube root of the final state mean particle multiplicity $\langle dN_{ch}/d\eta \rangle^{1/3}$~\cite{Alice2011,alice276,ALICE:2010igk,ALICE:2015hav,ALICE:2011kmy,ALICE:2015tra,ALICE:2020mkb}. A similar scaling across collision energies and colliding systems is only approximate~\cite{sinakk2004,shapoval2013}. 
The radii versus the pair transverse momentum exhibit a power-law-like scaling in pair transverse mass $m_{T}$ ($m_{T}$ = $\sqrt{k_{T}^2+m^{2}}$, where $k_{T}$ is pair mean transverse momentum, $k_{T}=|\textbf{p}_{T1}+\textbf{p}_{T2}|$/2, and $m$ is particle mass) \cite{STAR2001,PHENIX2004,STAR2005,Lisa2005,star2009,Alice2011,alice276,alice2020}.  
In heavy-ion collisions both these scalings are predicted by hydrodynamic models~\cite{yu2015, Kisiel:2014upa}. %i added therminator citation.. i guess because was missing this referee was doubting
%~\cite{sinyukov3,ALICE:2015tra}.  
Specifically, the $m_{T}$ scaling is explained as a direct consequence and one of the main signatures of the collective radial flow of the system.
%There were also observed a linear scaling of radii drawn in function of a final state multiplicity $\langle dN_{ch}/d\eta \rangle^{1/3}$ (at near the same initial sizes of creating  systems \cite{sinyukov3} ) as well as a power-law scaling on $m_{T}$ ($m_{T}$ = $\sqrt{k_{T}^2+m^{2}}$, where $k_{T}$ is an average pair transverse momentum $k_{T}=|\textbf{p}_{T1}+\textbf{p}_{T2}|$/2, and $m$ is a mass of particle). See more in e.g. \cite{STAR2001, PHENIX2004, STAR2005, Lisa2005, star2009, Alice2011, alice276, alice2020}. 

%Therefore the source radii are extracted from the Gaussian fits to the measured two-particle correlation functions at given pair momentum

%The spatio-temporal source is summarized by its shape and sizes
%The homogenity region is determined by interferometry radii that expresses it's sizes in three dimensions, the shape and orientation of the emission[][] as well as the dynamics of the collision processes [][]. 

The observed $m_{T}$ power-law scaling means %is reflected in measurements by 
a decrease of the radii with growing $m_{T}$ that is a signature of hydrodynamic collectivity, typical for all particle species affected by the same flow field. Such dependence can be helpful for the prediction of the source size at certain $m_{T}$ through the interpolation between radii measured at different $m_{T}$ values. Nonetheless, the $m_{T}$ range of experimental radii measurement depends on the acceptance of the detectors used in the study, and the very-low-$m_{T}$ region has not been reached yet via such measurements. 
Therefore, the femto-scale predictions in that region can come only from the extrapolation, which following the commonly used power-law function, shows a very rapid growth of the radii with decreasing $m_{T}$. 
The theoretical studies,
by contrast, do not face such problems and so, simulations within
a realistic collision model can help estimate the possible character
of the femtoscopy radii behaviour in ultra-soft momentum region.

The knowledge of femtoscopic radii for ultra-soft pions is becoming increasingly important in the context of rapidly developing experimental analyses of femtoscopic correlations of non-identical particles~\cite{Lednicky:2005tb,Kisiel:2009eh}. In the case of such pairs the correlation occurs between two particles with similar velocity. If the pair in question contains a pion and a relatively massive particle (e.g. a deuteron or a charmed meson), then the pion in the pair needs to be ultra-soft in order to be correlated. The femtoscopic radius of the pair is directly related to the size of emission regions of both particles in the pair. Therefore, in order to study the non-identical correlations between pions and massive particles, the knowledge of femtoscopic radii of ultra-soft pions becomes essential.

Several previous theoretical studies of the ultra-low momentum pions were motivated by the experimental evidence of enhanced production~\cite{ALICE:2012ovd,ALICE:2013mez}. These observations could be addressed by non-equilibrium models~\cite{Begun:2013nga}. 
In addition to the semi-classical models, pure quantum effects such as pion condensation and Bose-Einstein enhancement were also explored~\cite{Begun:2015ifa}. 
However, none of these studies addressed the pion femtoscopic radii nor their scaling at very low transverse momenta. 

Thus, in this paper, we focus on studying the pion femtoscopic 
%$\pi^{ch}$-$\pi^{ch}$ 
radii dependence on pair $m_{T}$, including the region of very low transverse masses, in Pb-Pb collisions at the LHC energy
$\sqrt{s_{NN}}$ = 5.02 TeV simulated within the integrated HydroKinetic Model (iHKM)~\cite{yu2015,yu2016,yu2019} and the hydrodynamics model~\cite{bozek2012} coupled to the statistical hadronisation code THERMINATOR2  (LHYQUID + THERMINATOR2)~\cite{CHOJNACKI2012746} --- LQTH.

\section{\label{sec:level1} MODELS' DESCRIPTION}
% SIMULATION DATA

The collisions of heavy ions at ultra-relativistic energies allow to create  quark-gluon plasma (QGP) --- a new state of strongly interacting matter, where quarks and gluons are no longer confined within individual nucleons. The evolution of such matter can be 
%more or less 
successfully described in complicated models realized in the form of codes for labor-consuming computer simulations.
Such codes typically include modules describing the initial state formation, relativistic hydrodynamics expansion of liquid-like QGP at the intermediate stage of the system's evolution, its subsequent particlization (turning into a set of hadrons) and, finally, the hadron-resonance gas expansion.
As a result of typical calculation, one obtains from the model a set of created hadrons, characterized by the space-time points of their last collision and 4-momenta. Based on these data one can construct different observables, like spectra, correlation functions etc.
%codes adapted to heavy-ion collisions at the Large Hadron Collider (LHC) at CERN. The source radii extracted from correlations describe the system at kinetic freeze-out, i.e., the last stage of particle interactions.

The data analyzed in this study were generated using two such models, 
iHKM and LQTH, each having its own characteristics.
%peculiarities.
%particularly, in the implementation of freeze-out process
%, called in this paper as iHKM and LHYQUID+THERMINATOR2. 

The iHKM is one of the most complete models, describing all the essential
phases of the matter evolution in the course of a relativistic A+A collision (see~\cite{yu2015,yu2016} for details).
The initial conditions (IC) for each simulation include the energy density spatial distribution at the starting time $\tau_0$, usually close to 0.1~fm/$c$ (for the high-energy collisions we use GLISSANDO code~\cite{gliss} to generate it), and anisotropic momentum distribution, inspired by the Color Glass Condensate model. These IC correspond to the very initial non-equilibrium partonic state right after the two nuclei collision. At the next stage of the system's evolution, it gradually thermalizes and approaches to
a nearly hydrodynamical state. Then (starting from the thermalization 
time $\tau_{th}\approx1$~fm/$c$) follows a continuous medium expansion described in the Israel-Stewart viscous hydrodynamics formalism.
As the matter expands and cools down, it eventually reaches the ``particlization temperature'' $T_p \approx 160$~MeV (depending on the
QGP equation of state used at the hydrodynamics stage), when one switches
to the description of the system in terms of hadrons (here the Cornelius routine~\cite{cornelius} is applied). Produced hadrons are then fed to the UrQMD hadron cascade code~\cite{urqmd1998,urqmd1999}, performing multiple hadronic re-scatterings and resonance decays
taking place at this final ``afterburner'' phase of the collision.
The model is calibrated based on the experimental mean charged particle multiplicity and pion $p_T$ spectrum slope in the most central collisions
of a given type. In this paper, we use the iHKM tuning, used for the LHC
$5.02 A$~TeV Pb-Pb collisions simulation described 
in detail in~\cite{yu2019}.

The LQTH is similarly properly calibrated to describe the $\sqrt{s_{NN}}=5.02$~TeV Pb-Pb collisions at the LHC. The model assumes a second order viscous (3+1)D hydrodynamics evolution of the created fireball including shear and bulk viscosity as well as the Israel-Stewart stress corrections. The hydrodynamic expansion starts at 0.6~fm/c and lasts until the single (both chemical and kinetic) freeze-out at a selected temperature ($T_p\approx140$~MeV), where statistical hadronization takes place. The system's chemical composition and momentum spectra do not change anymore, except for the resonance propagation and decays, which are simulated by the THERMINATOR2 package. At the freeze-out temperature, particles are generated according to statistical rules from the freeze-out hypersurface following the Cooper-Frye formula. The model reproduces the heavy-ion collisions results including flow and femtoscopic measurements ~\cite{bozek2012, therm2014,therm2018}. All details about the model implementation and initial conditions can be found in ~\cite{bozek2012}.

%The LQTH approach assumes the ideal (3+1)D hydrodynamics evolution of the created fireball (however, bulk and shear viscosity can be implemented as well) until the single (both chemical and kinetic) freeze-out at certain temperature ($T_p\approx140$~MeV), where statistical hadronization takes place, and the system's chemical composition and momentum spectra do not change anymore, except for the resonance decays. \textcolor{blue}{More details about model implementation and initial conditions can be found in ~\cite{bozek2012}.} As well as the iHKM, for the current study the model is properly calibrated to describe the $\sqrt{s_{NN}}=5.02$~TeV Pb-Pb collisions at the LHC.

\begin{figure}
\includegraphics[width=0.99\linewidth]{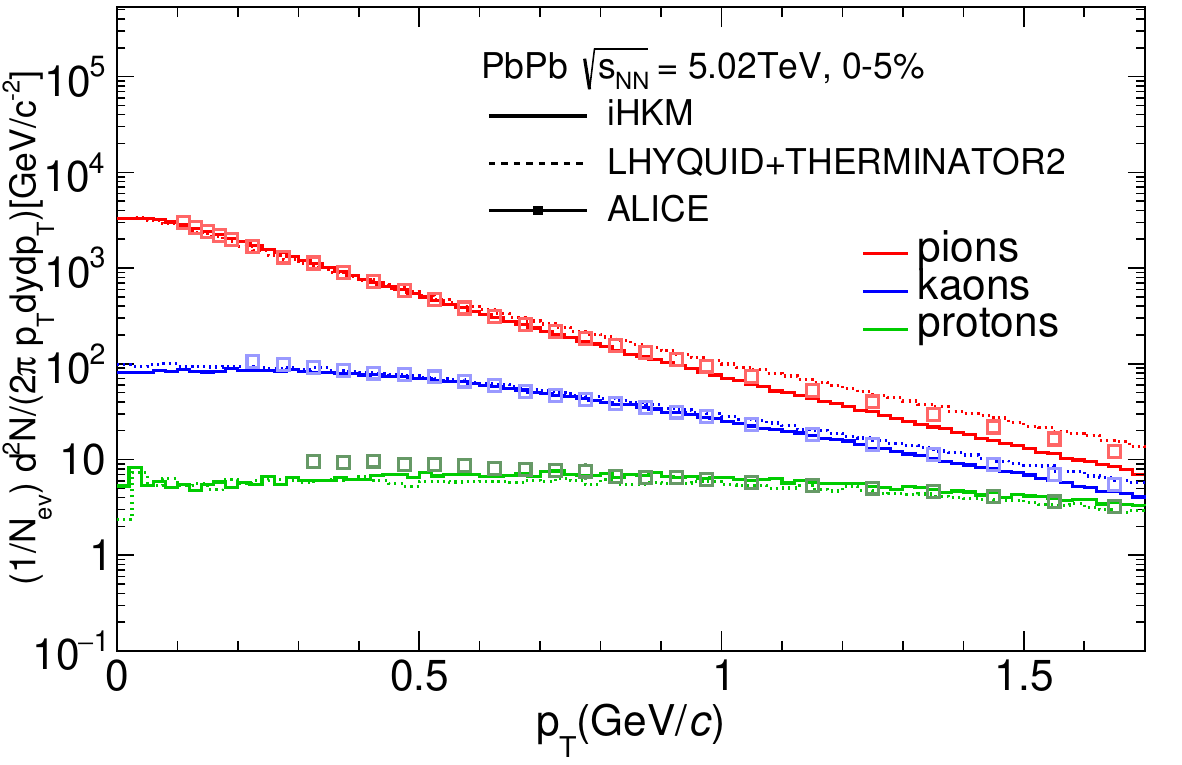}
\caption{\label{fig:pt} Pion, kaon
and proton transverse momentum spectra calculated in the iHKM (solid lines)
and in the LQTH models (dashed lines) for the most central $\sqrt{s_{NN}}=5.02$~TeV Pb-Pb collisions at the LHC~\cite{alicepsectra}.
}
\end{figure}

Both models describe well the experimental results for pion, kaon
and proton transverse momentum spectra, including the region of
very soft momenta (see Fig.~\ref{fig:pt}). Note, that such a description
is achieved without the implementation of additional specific mechanisms of
the soft pion emission, like the Bose-Einstein condensation, discussed,
e.g., in~\cite{Begun:2015ifa}. Also, the two utilized models were previously used to successfully describe the LHC data on $2.76 A$~TeV Pb-Pb collisions~\cite{yu2016,yu2019,therm2014,therm2018}. 

\section{\label{sec:level2} CONSTRUCTION OF THE CORRELATION FUNCTION}
The femtoscopic radii constituting the main subject of this study are typically defined from the Gaussian fits to the correlation functions depending on pair relative momentum $q=p_{1}-p_{2}$. 
%The space-time scales of the system are studied by the $\pi^{\pm}$-$\pi^{\pm}$ 
Here we consider the correlations in pairs of identical charged pions ($\pi^{+}$ and $\pi^{-}$), performing 1D and 3D analysis. The 1D correlation of identical particles is calculated as a function of $q_\mathrm{inv}=\sqrt{q_0^2-\textbf{q}^2}$ in the Pair Rest Frame (PRF) and requires minimal statistics in contrast to the 3D study. However, the radius $R_\mathrm{inv}$, in this case, is the only length extracted, so that the emission source is assumed to be a spherically symmetric Gaussian one. The 3D study is based on relative momentum components $q_\mathrm{out}$, $q_\mathrm{side}$, and $q_\mathrm{long}$ calculated in the Longitudinally Co-Moving System (LCMS), where the longitudinal direction is along the beam axis, the outward direction is along the pair transverse momentum and the sideward direction is perpendicular to the other two. Accordingly, the three radii $R_\mathrm{out}$, $R_\mathrm{side}$, and $R_\mathrm{long}$ are extracted from the CF fit in this case.

The correlation function is generally defined as %by Equation ($\ref{eq:1}$)
\begin{equation}
\label{eq:1}
    C(p_{1},p_{2})=\frac{P_{12}(p_{1},p_{2})}{P_{1}(p_{1})P_{2}(p_{2})},
\end{equation}
and can be understood as the ratio of conditional probability to detect a pair of particles with the specific momentum values, $P_{12}(p_{1},p_{2})$, %divided by 
to the probability of finding them with such momenta independently, $P_{1}(p_{1}) P_{2}(p_{2})$. 

In experimental analysis, the procedure of obtaining the correlation function is based on making pairs of detected particles. One uses particles coming from the same event to build the correlated distribution (numerator), while for the background distribution (denominator) pairs are created using particles coming from different events (meaning that they retain all aspects of experimental acceptance, while not being correlated due to the mutual interaction). 

In simulation studies, one usually fills two histograms (1D or 3D, depending on the analysis type) representing the corresponding particle pairs'
relative momentum $q$ distribution --- one histogram for the numerator and another one for the
denominator in Eq.~(\ref{eq:1}) --- based on the generated model output (set of hadrons). However, since in all currently available heavy-ion collision models the quantum statistical effects (including the Bose-Einstein symmetrization of the two identical bosons' wave-function) are not  implemented directly, on a microscopic level, one usually
%. We overcome this problem in the standard way by introducing 
introduces a weight factor
\begin{equation}
\label{eq:5}
    w = 1+\cos{(q r)},
\end{equation} 
coming from the Bose-Einstein interference at the so-called ``afterburner'' stage (to simplify the presentation, Coulomb and strong final-state interactions are not considered here). This factor $w$ is taken as a weight for the pairs entering the correlated distribution histogram (a numerator one). For the background distribution (denominator) the weight is equal to 1.
In this study, we also use the rapidity cut, $|y|<1$, during the identical charged pion pairs selection to reproduce the acceptance of the ALICE detector.

To obtain a fitting formula for the CF, which will allow us to define
the desired femtoscopy radii, one can consider the 7D (if particles are on the mass shell) particle emission functions $S_i(x,p)$ for each particle species $i$ and derive the approximate expressions for single-particle
and two-particle momentum spectra, and then using Eq.~(\ref{eq:1}) obtain the well known Bertsch-Pratt~\cite{pratt:1986, Bertsch:1989} representation for the correlation function of two identical bosons in LCMS: 
\begin{eqnarray}
\label{eq:7}
    C(\textbf{k,q})=1+\lambda_{3D}(\textbf{k}) \, \mathrm{exp}(-R^{2}_\mathrm{out}(\textbf{k})q^{2}_\mathrm{out}-\nonumber\\
    R^{2}_\mathrm{side}(\textbf{k})q^{2}_\mathrm{side}-R^{2}_\mathrm{long}(\textbf{k})q^{2}_\mathrm{long}),
    %_{out},q_{side},q_{long}
\end{eqnarray}
where femto-radii $R_i$ depend on mean pair momenta $\textbf{k}$, 
and $\lambda_{3D}$ are referred to as correlation strength factors.

%We model  particle emission in heavy ion collision in detail, therefore have full picture of the particle radiation at the collision processes, namely, we are able consider the 7-dim (if particles are on the mass shell) emission function $S_i(x,p)$ for all particle species $i$. It allows us to use formalism of emission function and then write correlation function in LCMS through  

%For comparison of our results with the formula~(\ref{eq:7}) one should take into account that all 
 
The invariant one-dimensional form of the correlation function Gaussian
parameterization is % well-known and looks as follows,  
\begin{equation}
\label{eq:6}
    C(k,q_\mathrm{inv})=1+\lambda_{1D}(k) \exp{(-R^{2}_\mathrm{inv}(k)q^{2}_\mathrm{inv})}.
\end{equation}

%SOURCE SIZE OF PIONS
\section{\label{sec:level3} PION FEMTOSCOPY SCALES}
This section presents the results on pion femto-radii obtained as a result of correlation function fitting for identical pions in 1D and 3D studies using the two models (iHKM and LQTH). The correlation functions were calculated separately for the six centrality classes ($0-5\%, 5-10\%,10-20\%, 20-30\%, 30-40\% $ and $40-50\%$) and the ten $k_{T}$ ranges $(0.00-0.05,0.05-0.10,0.10-0.15,0.15-0.20,0.20-0.25,0.25-0.30,0.30-0.35,0.35-0.45,0.45-0.60,0.60-0.80)$~GeV/$c$. The results from both models were based on approximately 2$\cdot 10^5$ events, and the statistical uncertainty of each correlation function was much smaller than the systematic effects.
%The number of events were selected 

\begin{figure}
\includegraphics[width=0.94\linewidth]{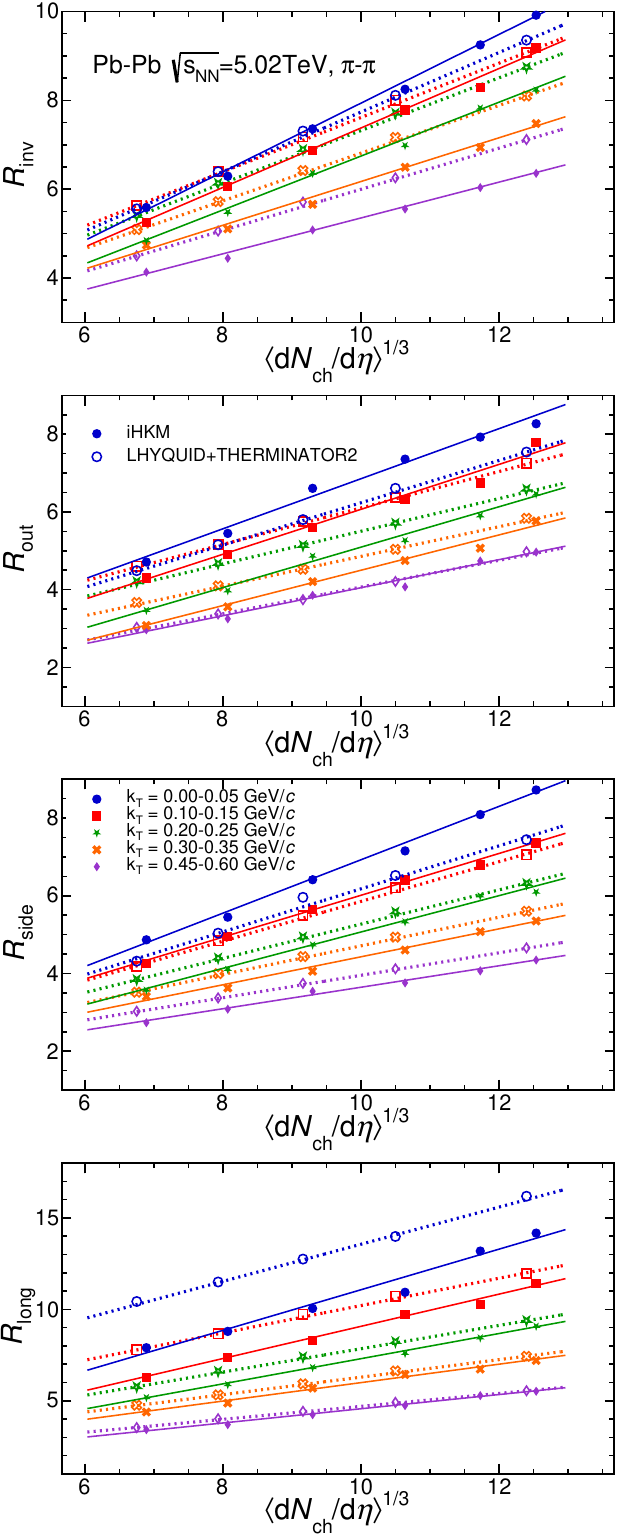}
\caption{\label{fig:wide} Femtoscopic radii of %$\pi^{\pm}$-$\pi^{\pm}$ 
charged identical pions as functions of \mbox{$\langle dN_{ch}/d\eta\rangle^{1/3}$} calculated using iHKM (full markers) and LQTH (empty markers) models, for 5 $k_{T}$ ranges (from top to bottom: $0.00-0.05,0.10-0.15,0.20-0.25,0.30-0.35,0.45-0.60$~GeV/$c$). Different panels from top to bottom correspond to 1D $R_\mathrm{inv}$, and 3D $R_\mathrm{out}$, $R_\mathrm{side}$, and $R_\mathrm{long}$ radii respectively. Lines correspond to linear fits to the radii dependencies. Symbols are slightly shifted in the x direction for visibility.}
\end{figure}

\begin{figure}
\includegraphics[width=0.96\linewidth]{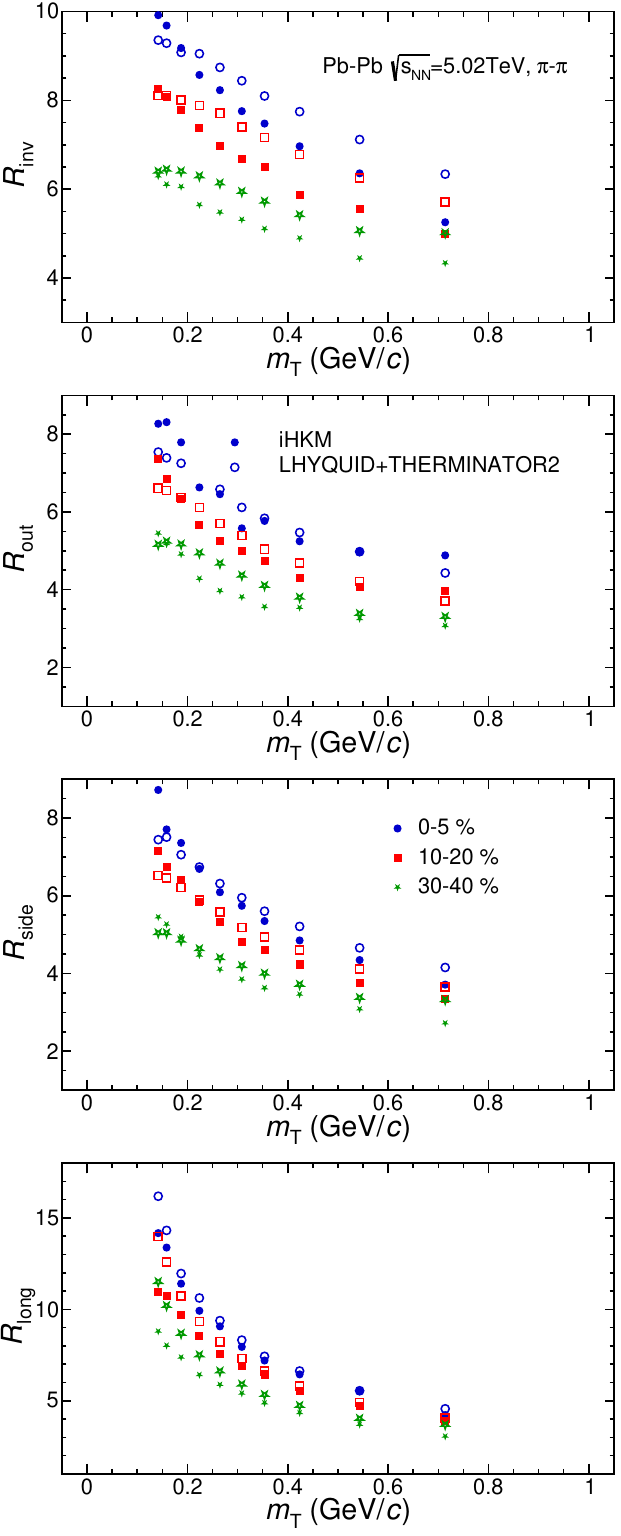}
\caption{\label{fig:wide2} Femtoscopic radii of %$\pi^{\pm}$-$\pi^{\pm}$ in 
charged identical pions as functions of $m_{T}$ calculated using iHKM (full markers) and LQTH (empty markers) models, for the three centrality classes ($0-5\%$ --- blue, $10-20\%$ --- red, $30-40\%$ --- green). Different panels from top to bottom correspond to 1D $R_\mathrm{inv}$, and 3D $R_\mathrm{out}$, $R_\mathrm{side}$, and $R_\mathrm{long}$ radii respectively. }
\end{figure}

\begin{figure*}
\includegraphics[width=0.95\linewidth]{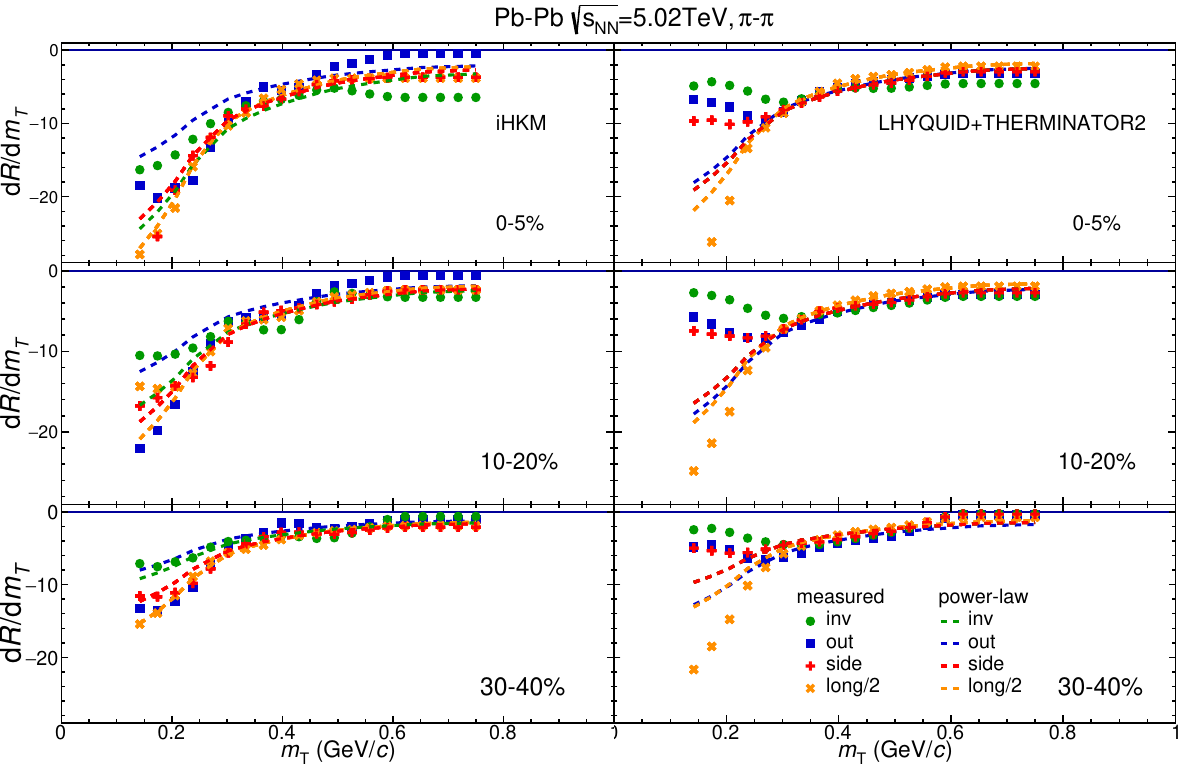}
\caption{\label{fig:deer} First-order derivative of the  
%of $\pi^{\pm}$-$\pi^{\pm}$ 
charged identical pions femtoscopic radii $m_{T}$ dependency, $dR/dm_T$, in the HKM (left) and the LQTH (right) models, for the three collision centralities (from top to bottom: $0-5\%, 10-20\%, 30-40\%$). Each panel contains derivatives of the three radii from the 3D study ($R_\mathrm{out}$ --- blue, $R_\mathrm{side}$ --- red, $R_\mathrm{long}$ --- orange) and $R_\mathrm{inv}$ from the 1D study (green). Markers represent results obtained from simulations and the dashed lines are obtained for the expected power-low trend based on fits to simulation points above 250~MeV/$c$.}
\end{figure*}

In Fig.$\ref{fig:wide}$ all the radii for identical pion pairs, corresponding to 1D and 3D calculations in the two models are
demonstrated as functions of \mbox{$\langle dN_{ch}/d\eta\rangle^{1/3}$}.
The results for the five of all ten $m_{T}$ bins are shown for better visibility since the radii in all bins follow a monotonic trend. The 
radii values vary in the range from about 15~fm ($R_\mathrm{long}$
for the lowest $m_{T}$ and the highest multiplicity in LQTH)
to about 2~fm ($R_\mathrm{side}$ for high $m_{T}$ and low multiplicity in iHKM).
%biggest sizes, reached in LHYQUID+THERMINATOR2 model for the lowest $m_{T}$ and the highest multiplicity events, are of the order of 15~fm in the longitudinal LCMS direction and around 9~fm in transverse directions of 3D study and radii of 1D study. The smallest measured radii come from the iHKM model for high $m_{T}$ and the lowest multiplicity events and are of the order of 2~fm. 
All the radii universally grow linearly with the cube root of final-state multiplicity. The multiplicity dependencies for various $m_{T}$ ranges generally show $m_{T}$ ordering, except for 1D and transverse 3D results from the LQTH model, where the first bins of $m_T$ overlap. 
%(the distance between all full markers and high $m_{T}$ open markers is bigger than between low $m_{T}$ open markers). 
This is because in the LQTH model, radii of the first $m_{T}$ bins take similar values.  Fig.~\ref{fig:wide2}, where femto-radii are shown as functions of $m_{T}$ presents it more clearly. All the measured radii decrease with increasing $m_{T}$. However, in the case of the LQTH model, we can observe a small plateau for the 1D and the transverse 3D radii at low $m_{T}$. The iHKM results, however, mostly fall monotonically in the lowest $m_{T}$ region. The Figure~\ref{fig:wide2} also exhibits an expected ordering with respect to analysed centrality --- the more peripheral collision, the smaller the radii. In both figures \ref{fig:wide} and \ref{fig:wide2}, the difference in femto-scales between the two models is also evident. 

One can see from Fig.~{\ref{fig:wide2} that the femto-radii in the iHKM model decrease faster than in the LQTH one. This is likely due to a very early, at $\tau_0 = 0.1$~fm/$c$, expansion starts in the iHKM,  leading to stronger velocity gradients at the final stage. This is the main reason for the reduction of the femto-radii: the homogeneity lengths are reduced with the growing flow intensity. So, at the intermediate and high $k_T$ the iHKM radii are smaller, than those in the LQTH model with a later start of the expansion ($\tau_0\approx0.6$~fm/$c$). As for the radii values at ultra-soft transverse momenta, the situation becomes even more complicated. %As one can see from 
%Fig.~{\ref{fig:wide2}, 
At very small $k_T$ the transverse radii are larger in the iHKM approach (the mentioned gradient could be the reason), while the longitudinal ones are larger in the LQTH model. %It means that 
The physics of pion production in the very-low-$k_T$ region appears to be non-trivial and worthy of further study and better understanding.

\begin{figure*}[tbh!]
\includegraphics[width=0.99\linewidth]{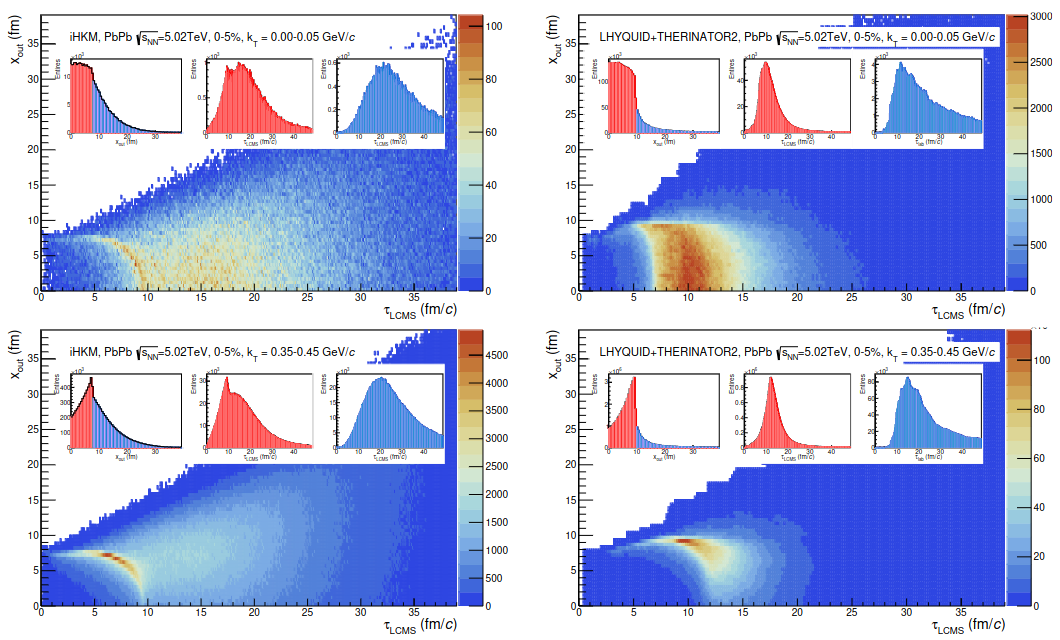}
\caption{\label{fig:dens} The pion last interaction time and transverse 
coordinate in out direction of LCMS system for iHKM (left) and LQTH (right) simulations of Pb-Pb $\sqrt{s_{\rm{NN}}}$=5.02 TeV collisions. The upper panels show distributions for the lowest $k_T$ interval, the bottom ones ---
for the middle $k_T$ interval considered in this study. Sub-figures show:  projection on $x_{out}$ direction (left), projection on time axis for the
two $x_{out}$ ranges (center and right).}
\end{figure*}

To better investigate the character of the radii $m_{\rm{T}}$ dependencies, they were differentiated, and the obtained derivatives are presented in Fig.~\ref{fig:deer} (for the three centrality classes). The derivative has been extracted in the range $\pm0.07$~GeV/$c$ around each point in the figure. %As it can be seen from the picture, 
Most of the functions show similar behaviour, with a clear change in the derivative %value 
form at certain momenta.
%changing at certain point. 
Namely, the derivative behaviour at low momenta ($m_{T}<0.25$~GeV/$c$), 
especially for the $R_{\rm{out}}$ and $R_{\rm{inv}}$ radii in LQTH model, is 
noticeably different from that of a power-law function derivative, 
typical for the higher momentum region and approximately followed 
even at low $m_{T}$ by the $dR_{
\rm{long}}/dm_{T}$ dependencies.
In more detail, in the iHKM one observes that the radius derivatives for 
the 1D case go flatter and are smaller by the absolute value than 
the $R_{\rm{long}}$ derivatives for all the collision centralities, while 
the \textit{out} and the \textit{side} curves go either slightly higher or 
slightly lower than the \textit{long} one, depending on the collision type.
In LQTH model we see a somewhat different situation, however,
rather similar for all the centrality classes: all the derivatives, 
except for the \textit{long} one, 
at low $m_{T}$ have much smaller absolute values than the power-law function
and demonstrate nearly flat behaviour with small falls and rises,
so that the radii dependency in this region can be sufficiently well approximated by a linear function.
%The iHKM model shows a constant derivative at low momentum (especially for 
%the 1D study), then a systematic rise and again a higher constant value at 
%large momenta. Flat parts of the curves correspond to constant slopes of the radii decrease approximated by a linear function. The rising/falling parts indicate a change in the radii dependency slope. 
%The low-$m_{T}$ flat part ends around 0.25~GeV/$c$. The LQTH model, in turn, shows for the 1D and the 
%3D transverse radii derivatives a constant or decreasing behaviour up to 0.3~GeV/$c$, then the dependence starts rising to end up with a constant value at large momentum. The initial decrease reflects the change of a nearly constant $R(m_T)$ dependence to a to decreasing one, while the next rising $dR/dm_T$ part is connected with more gentle radii change at growing $m_{T}$. The $long$ direction, by contrast, shows growing behaviour ending up with a plateau without any clear flat beginning.

\begin{figure*}
\includegraphics[width=0.95\linewidth]{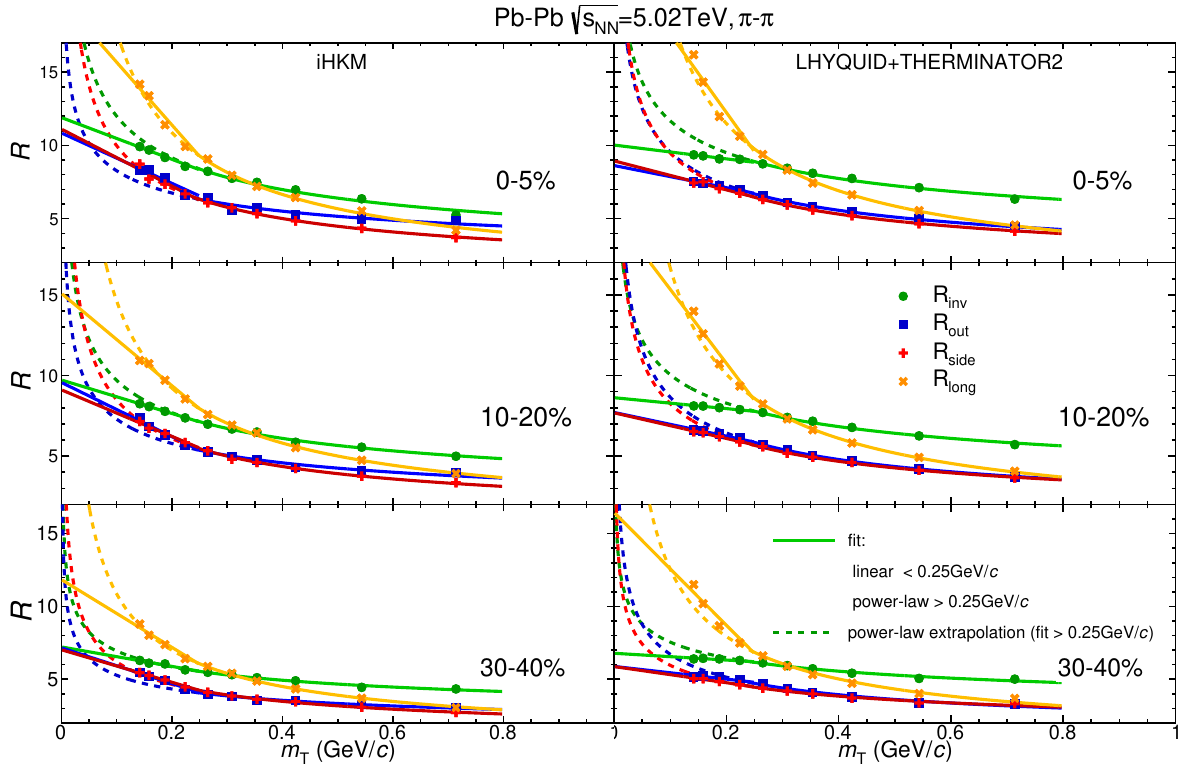}
\caption{\label{fig:fit} Pion femtoscopic radii dependencies on $m_{T}$ 
%$\pi^{\pm}$-$\pi^{\pm}$ in function of 
calculated in iHKM (left) and LQTH (right) models, for the three centrality classes (from top to bottom: $0-5\%, 10-20\%, 30-40\%$). 
Each panel contains results for the three radii from the 3D study ($R_\mathrm{out}$ --- squares, $R_\mathrm{side}$ --- crosses, $R_\mathrm{long}$ --- tilted crosses) and for $R_\mathrm{inv}$ from the 1D study (cicles). 
Lines correspond to power-law fits $a m_T^{-b}$ to radii dependencies in the range above 0.25~GeV/$c$.}
\end{figure*}

To interpret above mentioned results, one can use a very recent iHKM femto-analysis~\cite{Uni} as follows. When the system is just formed (at times near 0.1~fm/$c$), huge gradients of density in the transverse direction take place, since the system is essentially finite and borders with vacuum. The gradient is not equal locally along the radial directions: it is more strong at the periphery, and less strong in the center, where soft hadrons mostly come from. 
%As one can show~\cite{Uni}, 
In the vicinity of the geometrical center of the system, its decay into free particles happens at significantly later proper times than for the most of other parts of the system. Such a difference in proper times of spectra formation can be up to 5~fm/$c$~\cite{Uni}. Thus, in the context of the present study, one can expect that maximal formation times (about $15$~fm/$c$) should be typical for the ultra-soft pions with transverse momenta less than 0.3~GeV/$c$, that are emitted from the central region of the fireball. Whereas for the pions radiated from other parts of the decaying expanding system and having higher transverse momenta, $0.45-2$~GeV/$c$, the (proper) time of maximal emission is close to 10~fm/$c$~\cite{Uni}. The situation looks like a pion trap formed in the center of the created quark-gluon --- hadron system. The hadrons stay together (cannot leave the system) for a longer time in the system's center because of the following:

\begin{itemize}     
\item firstly, very low (close to zero) transverse collective velocity; 

\item secondly, smaller density gradient in the center during almost the entire duration of the evolution, as compared to non-central and periphery parts.

\item thirdly, the initially highest density in the geometrical center in central and semi-central nucleus-nucleus collisions. Therefore, for this high-mass-density region, it is difficult to expand because of the relatively small transverse pressure gradient in the center.
\end{itemize}

Note, that we see the mentioned effects of non-power-law transverse radii behaviour in the ultra-soft momentum region in ``hydro plus hadronic cascade'' models, where the dynamics is built in quasi-classical approximation. The Bose-Einstein quantum statistics effect is taken into account in the very final stage only, but even this way, using the femtoscopy method, one can see specific (delayed freeze-out) features of emission in the ultra-low-momentum region (see also Fig.~1 from~\cite{Uni}). 
In particular, as we already mentioned, the femtoscopy analysis brings us the signal about formation of a long-lived ultra-soft pion trap in the central part of the system created in ultra-relativistic heavy ion collision. 
To further check this signal, we built the distributions of the transverse \textit{out} coordinate of pion last collision points vs its time in the LCMS system, see Fig.~\ref{fig:dens}. It can be easily noticed that in both models the origin of particles forming low and 
high $k_{T}$ pairs is qualitatively different. 
Low $k_{T}$ pairs 
are mainly created of particles with low momenta most intensively emitted
from the decaying system close to its center. To see this, one can compare 
the red part of the $x_{\rm{out}}$ distributions shown in the panels of Fig.~\ref{fig:dens} for low (top panels) and high (bottom panels) particle momenta, respectively. The red parts correspond to the emission coming from the 
hadronizing hydrodynamic tube itself, while the blue parts correspond
to the emission at larger distances from the expanding hadron-resonance gas.
The blue parts are rather similar at low and high momenta, while the red ones
have a maximum in the center of the system for low momenta and at the periphery
of the system for higher momenta.
The red time distributions at low momenta have two maxima and larger mean
emission time than those at high momenta having only one maximum near
the system's particlization time. The blue time distributions are similar
for the low and high momentum cases.

The dependencies of the radii on $m_{T}$, together with the power-law fits, are presented for the three centrality classes in Fig.~\ref{fig:fit}. 
The fitting was performed outside the ``non-power-law'' region observed in Fig.~\ref{fig:deer} (above 0.25~GeV/$c$). 
As one can see from Fig.~\ref{fig:fit}, the power-law function in the form $a m_{\rm{T}}^{-b}$ works well for all the radii at not very low momenta, and in case of \textit{long} direction it describes even the ultra-soft region below 0.25~GeV/$c$.
Other radii, especially the 1D ones, at very low momenta go below 
(in the iHKM sometimes above) the power-law fitting curve.
%especially for all three $R_i$ in the 3D case in the iHKM model and for the $long$ direction of the LQTH model. As for the transverse radii from the LQTH, the first 2-3 corresponding points are mostly below expectations. In both models, the behaviour most different from the power-law in the low-$m_{T}$ region is observed for the 1D $R_\mathrm{inv}$ points, which lie, up to 0.25~GeV/$c$, below the extrapolated line. 
\begin{figure}
\includegraphics[width=0.85\linewidth]{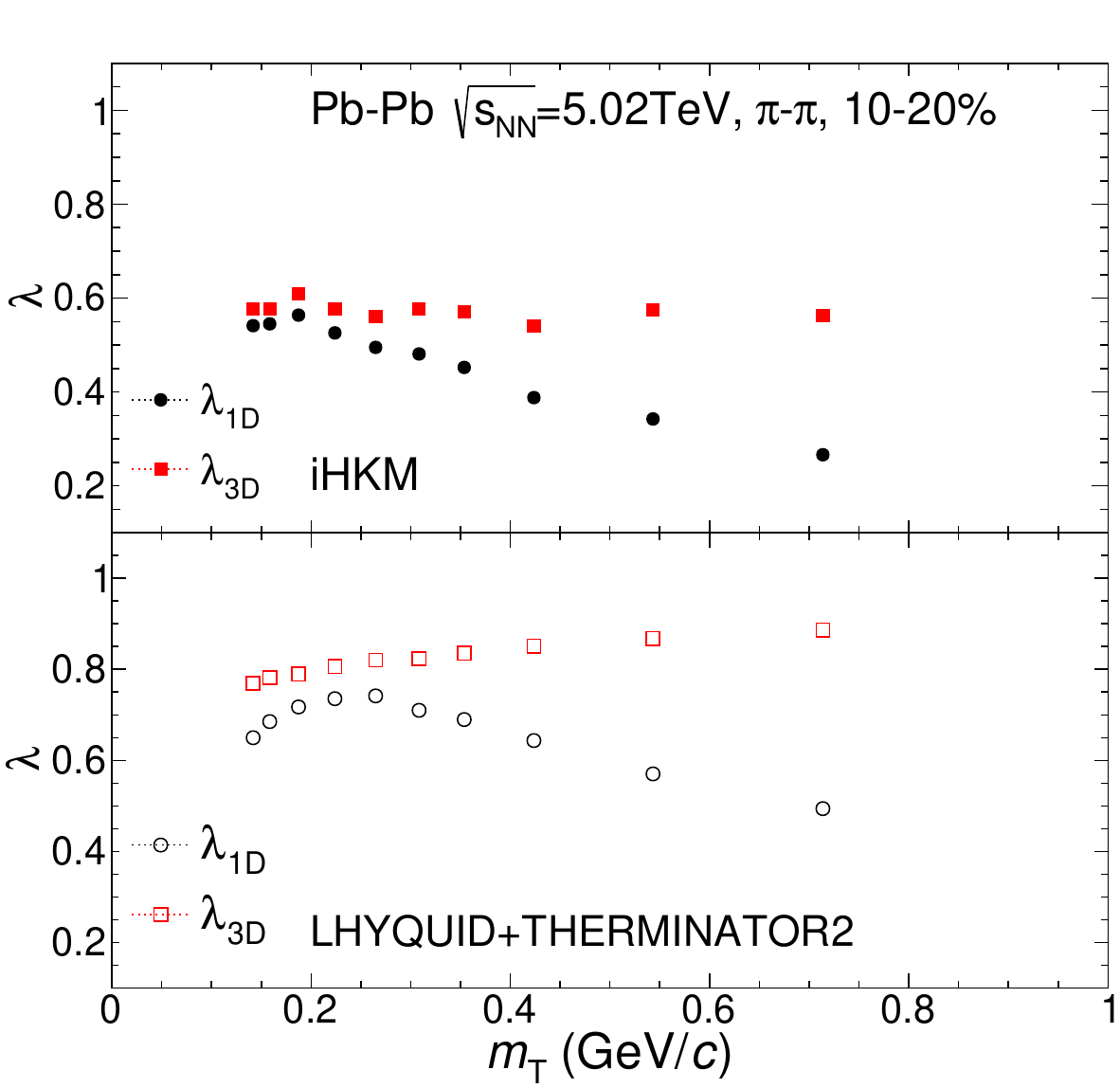}
\caption{\label{fig:lambda} The correlation strength $\lambda$ parameters from 1D (black) and 3D (red) fits to %$\pi^{\pm}$-$\pi^{\pm}$ 
the two-pion CF functions for the ten $m_{T}$ bins. 
The iHKM (top) and the LQTH (bottom) results are shown for $c=10-20\%$ Pb-Pb collisions at $\sqrt{s_{NN}}=5.02$~TeV. }
\end{figure}

In Fig.~\ref{fig:fit} the fall of 1D radii is less prominent than the fall for radii from 3D studies. 
%The low $m_{T}$ region seems to be a kind of average of three directions however, going further is bigger than the average. 
This can be connected with the shape of the respective correlation functions. The 1D functions are less Gaussian than the 3D functions, and therefore the fitting has to compensate for this by reducing the correlation strength parameter $\lambda$. The $m_T$ dependencies of lambda in the two applied models are shown in Fig.~\ref{fig:lambda}. In the low $m_{T}$ region the lambda behaves similarly in both cases, whereas with growing $m_{T}$ it tends to go down in the 1D case.

Finally, we analyzed the femto-radii dependence on pair velocity, see Fig.~\ref{fig:vel} for different centrality intervals and for both used model approaches. In all cases, we notice a slow radii falling at low velocities.
%that can indicate that velocity of particles is similar for first $k_T$ bins (?). 
All the distributions were fitted with the analytic formula derived on purpose of this study
%What we can propose as an explanation for this not working power-law in low $m_T$?\\
%something about alice run3 
\begin{equation}
    \label{eq:100}
    %C(q_{inv})
   R(\beta)=\frac{a}{e^{(\beta-b)/c}+1},
\end{equation}
where $\beta=k_T/m_T$ is pair velocity, and $a$, $b$, $c$ are free parameters of the fit. As particle velocity decreases, the radii seem to reach a saturated value, similar in a very broad range of $\beta$. Therefore such value could be used to determine an extrapolated value of the radii for ultra-soft pions, instead of the historically used dependence based on power-law $m_{T}$ scaling. It is particularly important for
non-identical femtoscopy studies, where one considers
pairs of particles with different masses moving at the same velocity  -- thus, having different momenta.
\begin{figure}
\includegraphics[width=0.85\linewidth]{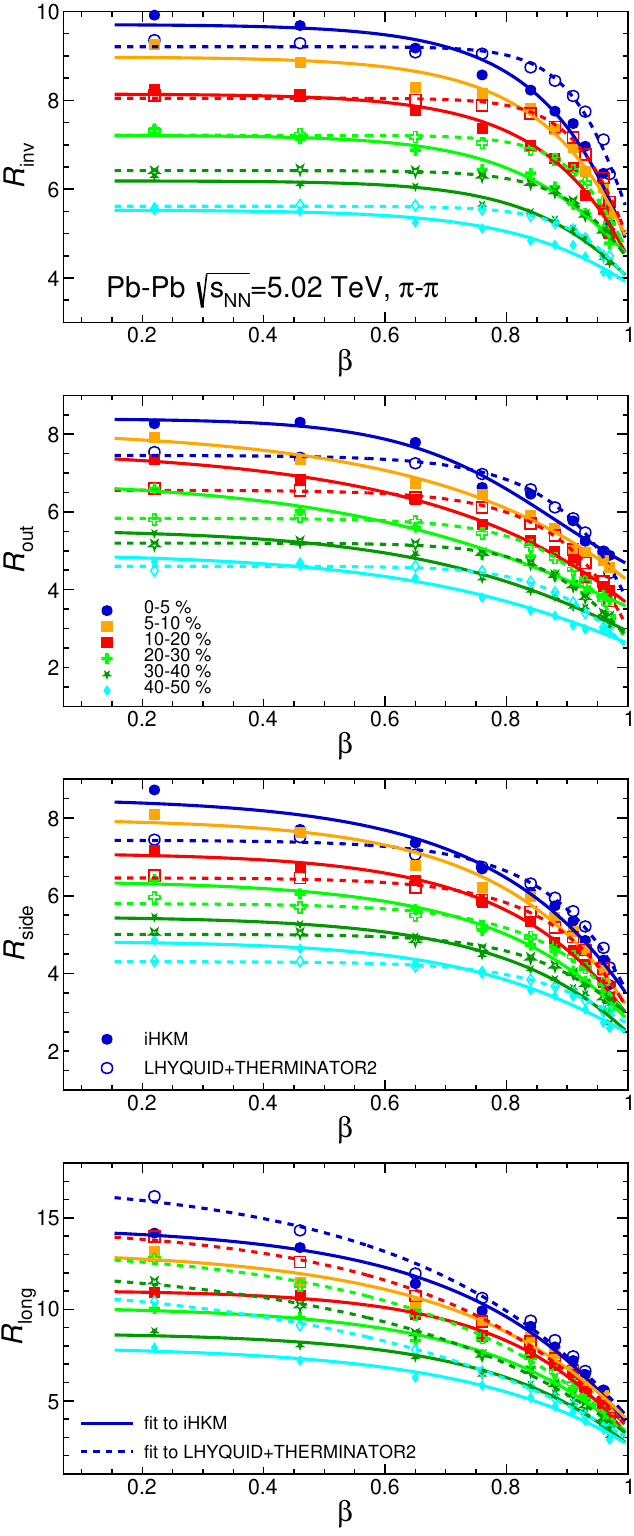}
\caption{\label{fig:vel} The identical pion femtoscopy radii dependencies on pair velocity $\beta$ in the iHKM (full markers) and the LQTH (empty markers) models. The results for the six centrality classes are shown. The lines represent fits to the radii by the formula (\ref{eq:100}).}
\end{figure}

\section{\label{sec:level4}SUMMARY}
In this paper, the model study of femtoscopic radii obtained from simulated data of iHKM and LQTH were shown. The results point out that the power-law character of the femtoscopic $m_T$-scaling is no longer valid for 1D studies in the low $m_{T}$ region (below 250~MeV$/c$). In the case of 3D studies, the radii are better described by the power-law dependence than 1D radii, however, the final conclusions are model-dependent. 
The iHKM model results are closer to a power-law behaviour, especially in the 3D analysis case.
%follows a power-law description in all 3 directions. 
On the other hand, for the LQTH model one sees noticeable deviations
from the power-law description everywhere, except for the longitudinal direction. Both models show that in low-$m_{\rm{T}}$ region radii distribution tends to be more linear than power-law or even flat in some cases. This results in smaller radii than expected from the extrapolation of the power-law scaling from intermediate and high $m_{\rm{T}}$.  

This modification of the behaviour is motivated by the capacity of the ultra-soft pions to decouple from the energetic and dense core of the expanding system, resulting in longer emission times but from the centre of the created system. 
This can be interpreted as both spatial and kinematic trapping of the ultra-soft pions by the more dense part of the created system. Such pions cannot escape this phase-space region until this confining mechanism is released. 
% relevance for non-id
This finding is fundamental for addressing experimentally the expected pion radii in non-identical femtoscopic studies when the second particle has a large mass difference with the pion mass, $e.g.$, pion--deuteron, pion--omega, or pion--charm hadron pairs. 

The observations of the non-monotonic behaviour of ultra-soft pions obtained from semi-classical models should also be explored from the quantum perspective. The role of confinement and its connection to the direct photon puzzle~\cite{gamma1,gamma2,gamma3,gamma4,gamma5} might as well influence the pion homogeneity lengths and should be studied further.

\section{\label{sec:level5}ACKNOWLEDGMENTS}
%I wanna thank me -- Snoop dog
%Last but not least, I wanna thank me
%I wanna thank me for believing in me
%I wanna thank me for doing all this hard work
%I wanna thank me for having no days off
%I wanna thank me for, for never quitting
%I wanna thank me for always being a giver
%And tryna give more than I recieve
%I wanna thank me for tryna do more right than wrong
%I wanna thank me for just being me at all times
This work is funded by the Research University – Excellence Initiative of Warsaw University of Technology via the strategic funds of the Priority Research Centre of High Energy Physics and Experimental Techniques, the IDUB POSTDOC programme, the IDUB YOUNG-PW programme, Scientific Council of the discipline grant programme and by the Polish National Science Centre under agreements 
no. 2022/45/B/ST2/02029, %Opus
%no. 2022/46/E/ST2/00255, %Sonata-bis
and no. 2023/49/N/ST2/03525% preludium 
.
The work was also supported by a grant from the Simons Foundation (Grant Number 1290596, Yu.S. and V.S.).

% The \nocite command causes all entries in a bibliography to be printed out
% whether or not they are actually referenced in the text. This is appropriate
% for the sample file to show the different styles of references, but authors
% most likely will not want to use it.
%\newpage
%\nocite{*}

\bibliography{bibliography}% Produces the bibliography via BibTeX.

\end{document}